\documentclass[aip,preprint,]{revtex4-1}
\usepackage{graphicx}
\usepackage{xcolor}
\usepackage{amssymb}
\usepackage{amsfonts}
\usepackage{amsmath}

\begin{document}

\title{Dynamical correlations and chimera-like states of nanoemitters
coupled to plasmon-polaritons in a lattice of conducting nanorings}

\author{Boris A. Malomed$^{1,2}$, Gennadiy Burlak$^{{\ast} 3}$, Gustavo Medina-%
\'{A}ngel$^{3,4}$, and Yuriy Karlovich$^{5~~}$}

\email{
gburlak@uaem.mx,malomed@tauex.tau.ac.il,gustavo.isc@hotmail.com,karlovich@uaem.mx
}

\affiliation{Department of Physical Electronics, School of Electrical Engineering,
			Faculty of Engineering, and Center for Light-Matter Interaction, Tel Aviv
			University, P.O.B. 39040, Tel Aviv, Israel}

\affiliation{Instituto de
Alta Investigaci\'{o}n, Universidad de Tarapac\'{a}, Casilla 7D, Arica,
Chile}

\affiliation{CIICAp, Universidad Aut\'{o}noma del Estado de Morelos, Av. Universidad 1001,
	Cuernavaca, Morelos 62209, M\'exico}

\affiliation{FCAeI, Universidad Aut\'{o}noma del Estado de Morelos, Av. Universidad 1001,
		Cuernavaca, Morelos 62209, M\'exico}
\affiliation{CInC, Universidad Aut\'{o}noma del Estado de Morelos, Av. Universidad 1001,
		Cuernavaca, Morelos 62209, M\'exico}

\begin{abstract}
We systematically investigate semiclassical dynamics of the optical field
produced by quantum nanoemitters (NEs) embedded in a periodic lattice of
conducting nanorings (NRs), in which plasmon polaritons (PPs) are excited.
The coupling between PPs and NEs through the radiated optical field leads to
establishment of a significant cross-correlation between NEs, so that their
internal dynamics (photocurrent affected by the laser irradiation) depends
on the NR's plasma frequency $\omega _{p}$.
The transition to this regime,
combined with the nonlinearity of the system, leads to a steep increase of
the photocurrent in the NEs, as well as to non-smooth (chimera-like or
chaotic) behavior in the critical (transition) region, where small
variations of $\omega _{p}$ lead to significant changes in the level of the
NE pairwise cross-correlations. The chimera-like state is realized as
coexistence of locally synchronized and desynchronized NE dynamical states.
A fit of the dependence of the critical current on $\omega _{p}$ is found,
being in agreement with results of numerical simulations. The critical
effect may help to design new optical devices, using dispersive nanolattices
which are made available by modern nanoelectronics.
\end{abstract}
\maketitle

\section{Introduction}

The current work in the field of infrared photoelectric information
technologies yields advances in the generation, manipulations, source
detection, and related achievements in the field of infrared photonics
\cite{Kun:2022,Perera:2011,Ravinder:2020, Dai:2019,Ryzhii:2020,Arrazola:2014}%
. The inclusion of dispersive nanorings (NRs) in the operation of
nanoemitters (NEs) considerably affects the generated electromagnetic field,
whose structure significantly depends on the NR's plasma frequency, $\omega
_{p}$ \cite{Guo:2018,Severin:2021a}. In such hybrid systems, it is possible
to control properties of local optical fields and the creation of
miniaturized low-threshold coherent tunable sources \cite%
{Severin:2021a,Burlak:2024}.
An essential feature of these structures in the
disordered state is that NE\ clusters produce fractal radiation patterns, in
which light is simultaneously emitted and scattered \cite{BURLAK:2023b}. In
the practically relevant case of lossy NRs with embedded NEs, other factors
become important too. Optical fields of dispersive NRs perturb the energy
levels of NEs, hence plasmon polaritons (PPs) populating NRs affect the
internal degrees of freedom of the quantum NEs coupled \ to the NRs.
Nonlinearity is an important feature of NEs, which leads to laser emission
\cite{Siegman:1986}. All that leads to resonant changes in the field
structure associated with the PP excitation in the NRs. We found that such a system
exhibits coexistence of locally synchronized and desynchronized dynamics of
random NEs, which may be considered as a chimera-like behavior in the
respective range of $\omega _{p}$. \textcolor{black}{We remind that a ``chimera state" is a
dynamical pattern that occurs in a network of coupled identical oscillators
when the oscillator population is broken into synchronous
and asynchronous parts \cite{Abrams:2004}} The PP field being external to the field
in the NEs, at small values of $\omega _{p}$ the NE\ dynamics is practically
independent of $\omega _{p}$. However, at overcritical values of $\omega _{p}
$, i.e., above the transition to the state with the strong coupling of the
NR and NE subsystems, the NE photocurrent essentially depends on $\omega _{p}
$.\

In this paper we
theoretically study dynamics and correlations of quantum NEs embedded in a
periodic lattice of conducting NRs, in which the PPs are excited. We show
that the coupling between PPs and NEs through the optical field leads to a
significant correlation between NEs, so that the internal dynamics of the
NEs (quantum photocurrent) depends on the plasma frequency $\omega _{p}$ of
the classical subsystem of PPs in the NR. Thus, the setup is built of two
subsystems coupled by the radiation field, \textit{viz}., the classical
array of the NRs and the quantum (actually, semi-classical) subsystem of
NEs. We consider conducting (carbon) NRs, whose properties are determined by
$\omega _{p}$. The PP field interacts with NEs and perturbs its quantum
degrees of freedom. \textcolor{black}{We show that the field dynamics is distinct in different ranges of
$\omega _{p}$, which can be separated by some characteristic value $\omega
_{c}$. At small $\omega _{p}<\omega _{c}$, the PP field has a small
amplitude, weakly perturbing the quantum dynamics and securing smooth
cross-correlations of the NEs. However, at larger $\omega _{p}\geq \omega
_{c}$, a transition occurs to the state in which the PP field produces a
significant contribution to the radiation, which perturbs the dynamics
of NEs, leading to a change in their cross-correlations.}

The rest of the paper is organized as follows. In Sec. 2, we formulate basic
equations for the considered hybrid NR-NE coupled system. In Sec. 3, we
study dispersion characteristics and the field structure of the plasmonic
modes in the NRs. In Sec. 4, we investigate the structure of the optical
field of the laser emission in the system and the PP-mediated dynamics of
the NEs coupled to NRs. In Section 5, we explore the phase transition
exhibited by the total PP current in the lattice. Section 6 concludes the
paper.

\begin{figure}[tbp]
\centering
\includegraphics[
width=0.6\textwidth
]{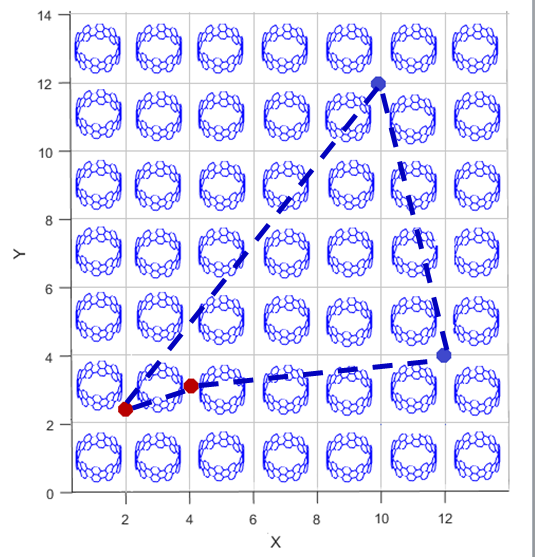}
\caption{The system is built as the lattice of size $7\times 7$. composed of
conducting NRs with embedded quantum NEs (solid circles) emitting the
optical field. In the present setup, four NEs form two clusters shown by red
and blue colors. PPs are excited in the NRs and interact with the NEs. The
NEs are connected by straight lines that correspond to the optimized path
calculated by means of TSP (traveling salesman problem) method (with a
traveling photon passing each NE without visiting the same NE twice) and
Fermat's principle\protect\cite{Burlak:2015,BURLAK:2023b}.}
\label{Fig_Latt_Emitts}
\end{figure}

\section{Basic equations}

The system under the consideration is Fig.~\ref{Fig_Latt_Emitts}) contains a
periodic 2D lattice of conducting NRs (NR) $(7\times 7)$ with a randomly
embedded rarefied set of quantum NEs (only four in Fig. \ref{Fig_Latt_Emitts}%
, shown by solid circles) emitting the optical field. PPs are excited in the
NRs and interact with the embedded NEs through the radiation field.

In Fig.~\ref{Fig_Latt_Emitts} the NEs are connected by straight lines
corresponding to the optimized path calculated using the TSP (traveling
salesman problem) technique and Fermat's principle \cite{Burlak:2015}. In
such a configuration,
the length of the connecting path $A_{D}$ is
proportional to the optimized (dimensionless) distance, which is a number of
nodes in the numerical grid, which a photon can travel in the sample,
passing each NE without visiting the same NE twice. Accordingly, $A_{D}$ is
calculated in the framework of our numerical analysis by dint of the TSP
(traveling salesman problem) technique \cite{Burlak:2015,BURLAK:2023b}. The
color scheme indicates that the embedded NEs in Fig.~\ref{Fig_Latt_Emitts}
consist of two clusters. To study this hybrid system, we use with
time-dependent Maxwell's equations in the lattice of 2D NRs, coupled to the
semi-classical rate equations for the electron populations in the NE \cite%
{Siegman:1986}.

The Maxwell equations are \cite{Soukoulis:2000,Burlak:2015}%
\begin{equation}
\nabla \times \mathbf{E}=-\mu _{0}\frac{\partial \mathbf{H}}{\partial t}%
\text{, }\nabla \times \mathbf{H}=\varepsilon _{0}\frac{\partial \mathbf{E}}{%
\partial t}+\mathbf{J}+\frac{\partial \mathbf{P}}{\partial t}\text{ \textbf{%
, } }  \label{MaxwellEqs}
\end{equation}%
were $\mathbf{J}=\sum_{k}{\mathbf{J_{k}}}(\mathbf{R}_{k}^{\mathrm{NR}%
},t)\delta _{\mathbf{rR}_{k}^{\mathrm{NR}}\text{ }}$ is the PP electrical
current in the NRs placed at spatial positions $\mathbf{R}_{k}^{\mathrm{NR}}$%
, and $\mathbf{P}=\sum_{k}{\mathbf{P_{k}}}(\mathbf{R}_{k}^{\mathrm{NE}%
},t)\delta _{\mathbf{rR}_{k}^{\mathrm{NE}}}$ is the electron polarization in
the embedded NE placed at $\mathbf{R}_{k}^{\mathrm{NE}}$. Here $\delta _{%
\mathbf{rR}}$ is the Kronecker's symbol, and the sums run over all NRs ($%
k=1,N_{r}$) and NEs ($k=1,N_{s}$). In Eq. (\ref{MaxwellEqs}) the electric
current of the conducting electrons in the NRs obeys the material equation
\cite{taflove:2005} $\mathbf{\dot{J}}_{k}+\gamma _{e}\mathbf{J}%
_{k}=\varepsilon _{0}\omega _{p}^{2}\mathbf{E}$, where $\gamma _{e}$ is the
collision frequency of electrons, $\omega _{p}$ is the plasma frequency, as
mentioned above, and $\varepsilon _{h}$ is the dielectric constant of the
host medium of NR. In the semi-classical approximation for non-interacting
electrons, the evolution equation for $\mathbf{P}_{k}$ in the vicinity of
the embedded NE is \cite{Siegman:1986}
\begin{equation}
\frac{\partial ^{2}\mathbf{P}_{k}}{\partial t^{2}}+\Delta \omega _{a}\frac{%
\partial \mathbf{P}_{k}}{\partial t}+{\omega _{a}^{2}}\mathbf{P}_{k}=\frac{%
6\pi \varepsilon _{0}c^{3}}{\tau _{21}\omega _{a}^{2}}(N_{1,k}-N_{2,k})%
\mathbf{E}_{k}.  \label{PolarizEq}
\end{equation}

\begin{figure}[tbp]
\centering
\includegraphics[
	width=0.50\textwidth
	]
	{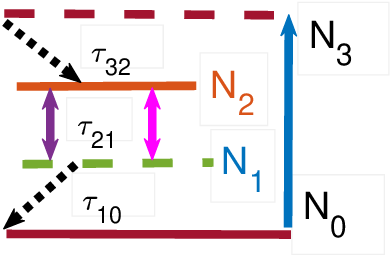}
\caption{The schematic representation of NE as a four-level system, see Eqs.
(\protect\ref{EqForN_03})-(\protect\ref{M_rateEq}). The external pump lifts
electrons from the ground level (with population $N_{0}$) to the third level
(with population $N_{3}$). After a short lifetime $\protect\tau _{32}$, the
electrons perform the nonradiative transfer to the second level (with
population $N_{2}$). The second level and the first level (with population $%
N_{1}$) are defined as the upper and lower lasing levels. Electrons are
transferred from the upper\ level to the lower one by both spontaneous and
stimulated emission. At last, electrons can perform the nonradiative
transfer from the first level back to the ground one. }
\label{Fig_4_level}
\end{figure}

To complete the model, we add the rate equations \cite{Siegman:1986} for the
occupation numbers of NEs, $N_{i,k}=N_{i}($\textbf{$\mathbf{R}_{k}^{\mathrm{%
NE}}$}$,t)$ (following \cite{Soukoulis:2000}, we assume that the NEs are
four-level quantum dots, as illustrated by Fig. \ref{Fig_4_level}):%
\begin{equation}
\frac{\partial N_{0,k}}{\partial t}=-A_{r}N_{0,k}+\frac{N_{1,k}}{\tau _{13}}%
\text{, }\frac{\partial N_{3,k}}{\partial t}=A_{r}N_{0,k}-\frac{N_{3,k}}{%
\tau _{02}},  \label{EqForN_03}
\end{equation}%
\begin{equation}
\frac{\partial N_{1,k}}{\partial t}=\frac{N_{2,k}}{\tau _{32}}-M_{k}-\frac{%
N_{1,k}}{\tau _{13}}\text{, }\frac{\partial N_{2,k}}{\partial t}=\frac{%
N_{1,k}}{\tau _{12}}+M_{k}-\frac{N_{2,k}}{\tau _{02}}\text{,}
\label{EqForN_12}
\end{equation}%
\begin{equation}
{M_{k}=\frac{(\mathbf{I_{p}}\cdot \mathbf{E})_{k}}{\hbar \omega _{a}}\text{,
}\mathbf{I_{p}}_{k}=\frac{\partial \mathbf{P}_{k}}{\partial t}}\text{.}
\label{M_rateEq}
\end{equation}%
Here $\Delta \omega _{a}=\tau _{21}^{-1}+2T_{2}^{-1}$, where $T_{2}$ is the
mean time between dephasing events, $\tau _{21}$ is the decay time for the
spontaneous transition from the second atomic level to the first one, $%
\omega _{a}$ is the radiation frequency (see e.g. \cite{Siegman:1986}), and $%
M_{k}$ is the induced radiation rate or excitation rate, depending on its
sign \cite{Soukoulis:2000}. Note that components $\mathbf{j}_{pk}$ parallel
to $\mathbf{E}_{k}$ mainly contribute to Eqs.(\ref{EqForN_12}) and (\ref%
{M_rateEq}) \cite{Downing:2020}. Coefficient $A_{r}$ is the pump rate for
the transition from the ground level ($N_{0}$) to the third one ($N_{3}$),
which is proportional to the pump intensity in the experiment \cite%
{Soukoulis:2000}.

\begin{figure}[tbp]
\centering
\includegraphics[
	width=0.84\textwidth
	]	{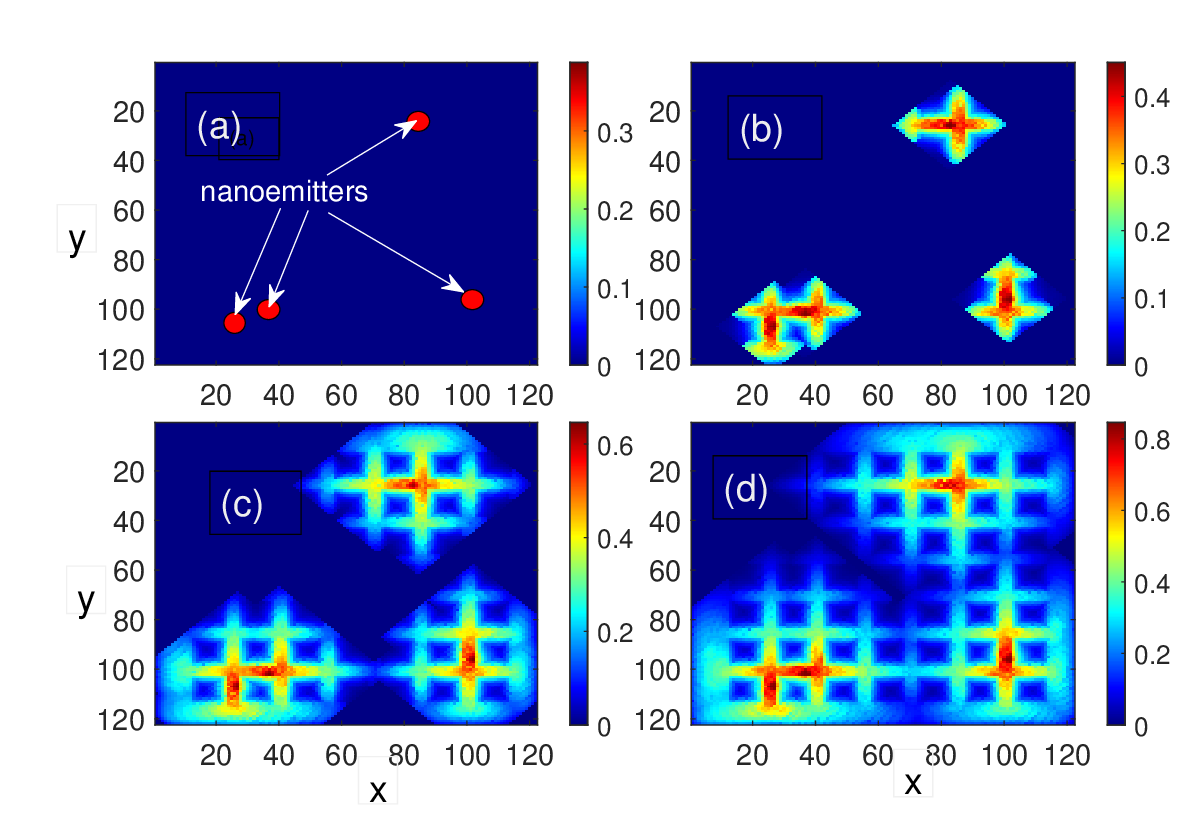}
\caption{The temporal dynamics (FDTD) of spatial distribution of the PP
field generated by NEs in the ${7}\times{7}$ lattice of conducting NRs (see
Fig.~\protect\ref{Fig_Latt_Emitts}) at $\protect\omega _{p}=2.3$~THz at
different simulating times $t_{f}$: (a) $20$, (b) $40$, (c) $60 $, (d) $80$.
The color coding represents values of the field amplitude.) At long times $%
t_f > 60$, the PP field, generated in the NR lattice by NEs, gradually
covers nearly the entire NR lattice. The local field (shown by the red
color) in the vicinity of NEs (see panel (a)) is the largest. The optical
field is concentrated in gaps of the NR lattice and practically does not
penetrate into the NRs. }
\label{Fig_NE energy}
\end{figure}

The finite-difference time-domain (FDTD) numerical method \cite%
{Taflove:2005a} was used to solve the model. In the simulations, we consider
the gain medium with parameters of the GaN powder, see Refs. \cite%
{Soukoulis:2000,Cao_exper:1999}. In Eqs. (\ref{PolarizEq}) - (\ref{EqForN_12}%
), frequency $\omega _{a}$ is $2\pi {\times }3{\times }10^{13}\,\mathrm{Hz}$%
, the lifetimes are $\tau _{32}=0.3\,\mathrm{ps}$, $\tau _{10}=1.6\,\mathrm{%
ps}$, $\tau _{21}=16.6\,\mathrm{ps}$, and the dephasing time is $%
T_{2}=0.0218\,\mathrm{ps}$. In what follows we use the dimensionless time $t$
normalized as $t\rightarrow tc/l_{0}$, where $l_{0}=100$ $\mathrm{\mu }${m}
is the typical spatial scale and $c$ is the light velocity in vacuum. Thus,
the present model couples the population-rate equations at different NE\
levels to the PP field equations in the vicinity of the NR lattice.
Therefore, the NE\ resonant emission operation in the system is affected by the PP\
excitation in the NRs, which finally leads to essentially nonlinear field
dynamics. 
Figure~\ref{Fig_NE energy} shows the temporal dynamics of the PP field
distribution in the ${7}\times {7}$ lattice of the conducting NRs (see Fig.~%
\ref{Fig_Latt_Emitts}), as produced by our FDTD simulations of Eqs. (\ref%
{MaxwellEqs})-(\ref{M_rateEq}) at $\omega _{p}=2.3$~THz at different
simulation times $t_{f}$. In Fig.~\ref{Fig_NE energy} the color-coding
scheme for the field amplitude shows that, at short times $<40$, the PP
field in the NR lattice is small. However, at longer times, an increasingly
stronger PP field is generated in the lattice, gradually covering the entire
lattice with time. The local field (designated by the red color) in the
vicinity of NEs is large, as expected. Figure.~\ref{Fig_NE energy}
illustrates the temporal dynamics of the nonlinear transition forming the
relationship between the NE oscillators and the PP in the underlying
lattice. It shows why the chimera-like states under the study (recall they
are called chimeras as they combine the synchronized and desynchronized NE
dynamical states) depend not only on time, but also on the plasma frequency $%
\omega _{p}$ of the surrounding NRs.

\section{Plasmon modes in the NRs (nanorings)}

The transmission characteristics of surface PPs in ring resonators have been
studied in nano-optics under the condition that the frequency dispersion may
be neglected\cite{Tong:2009,Wolff:1971}. However, this approximation is not
applicable to the conducting NR in a vicinity of $\omega =\omega _{p}$. In
the present section we briefly analyze the structure of the spectrum of a
single nanoparticle. Below, the optical field in the NR lattice is studied
by means of the FDTD technique, cf. Ref. \cite{Taflove:2005a}. As the most
fundamental object, we consider a single cylindrical nanoparticle of radius $%
R$ and length $L_{n}$, containing $N_{e}$ valence electrons, assuming that
the density of the valence electrons is uniform, $n(r)\equiv n_{0}$.
The cylindrical geometry admits the propagation of independent TM \textcolor{black}{($[E_\parallel,H_\perp]$) and TE
($[E_\perp,H_\parallel]$)} modes with the longitudinal and transverse components of the electric field,
the corresponding subscripts being denoted below as $\alpha =\parallel $ and
$\alpha =\perp $ components, respectively. Following \cite{Downing:2020},
we define shift $\mathbf{u}_{\alpha }$ of the electron distribution, hence
one can calculate the respective displaced density as $n(\mathbf{r}-\mathbf{u%
}_{\alpha })\approx n(\mathbf{r)+}\delta n_{a}$. For small $u_{\alpha }$,
one has
\begin{equation}
\delta n_{\alpha }(\mathbf{r})=-\mathbf{u}_{\alpha }\cdot \nabla n(\mathbf{r}%
).  \label{eq:Long_Hartree_delta}
\end{equation}%
The corresponding energy variation is
\begin{equation}
\delta {E}_{\alpha }=\frac{e^{2}}{2}\int \mathrm{d}^{3}\mathbf{r}\int
\mathrm{d}^{3}\mathbf{r^{\prime }}\,\frac{\delta n_{\alpha }(\mathbf{r}%
)\delta n_{\alpha }(\mathbf{r^{\prime }})}{|\mathbf{r}-\mathbf{r^{\prime }}|}%
,  \label{eq:Hartree_alpha}
\end{equation}%
therefore the restoring force is $F_{\alpha }=-\frac{\partial }{\partial {u}%
_{\alpha }}\left( \delta E_{\alpha }\right) =-k_{\alpha }u_{\alpha }$,
hence the normal-mode's frequency is
\begin{equation}
\omega _{0,\alpha }=\sqrt{\frac{k_{\alpha }}{M_{\mathrm{e}}}}.
\label{eq:normal_mode}
\end{equation}%
Here $M_{\mathrm{e}}=N_{\mathrm{e}}m_{\mathrm{e}}$ is the full mass of the
electrons, while $k_{\alpha }$ is the effective spring constant.

\begin{figure}[tbp]
\centering
\includegraphics[
	width=0.70\textwidth
	]	{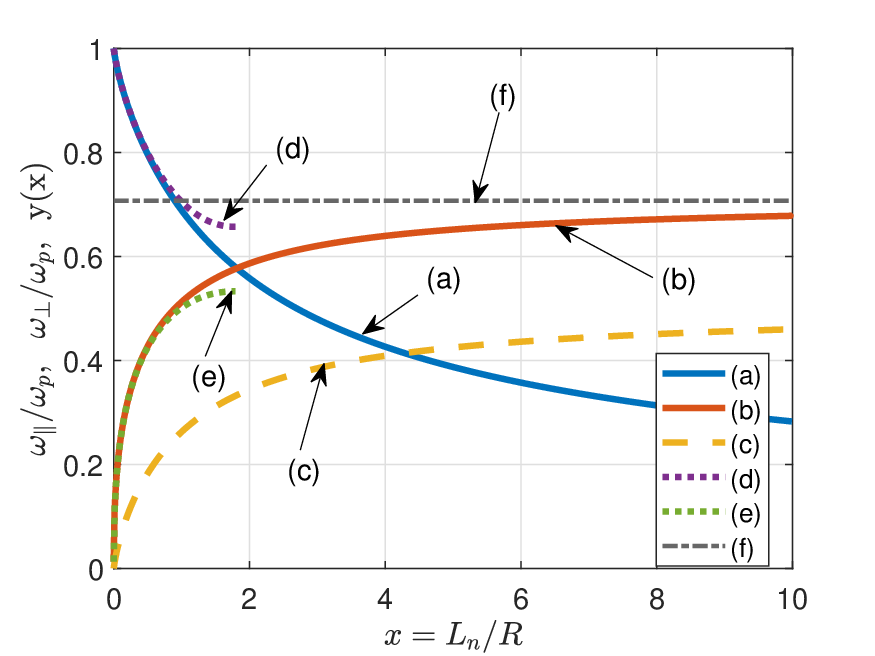}
\caption{Dependencies $\protect\omega _{\parallel }/\protect\omega _{p}$ (a)
and $\protect\omega _{\perp }/\protect\omega _{p}$ (b) on parameter $%
x=L_{n}/R$ in the interval $[0,10]$, see Eqs.(\protect\ref{w_parall}) and (%
\protect\ref{T_dir}) respectively. Dotted lines ((d) and (e)) show the
approximate solutions. It is seen that such an approximation is correct only
for $L_{n}/R<0.9$. At $x\approx 1.75$ the values of $\protect\omega %
_{\parallel }$ and $\protect\omega _{\perp }$ are close. The yellow dashed
line (c) displays the function $y(x)$ from Eq.(\protect\ref{L_dir}), and the
black line (f) displays the asymptotic value $1/\protect\sqrt{2}$ of the
normalized surface plasmon frequency \textcolor{black}{for $x \gg 1$}. }
\label{FigDow}
\end{figure}

For the longitudinal displacement $\mathbf{u}_{\alpha =\parallel }$, the
variation of the energy in Eq. (\ref{eq:Hartree_alpha}) can be written
as%
\begin{equation}
\delta E_{\parallel }=2\pi ^{2}\left( eun_{0}\right)
^{2}R^{3}x[1-2y(x)],~x=L_{n}/R,  \label{L_dir}
\end{equation}%
where $y(x)=(g(x)-4/3)/(\pi x)$, and
\begin{equation}
g(x)=\frac{x}{6}\left[ \left( x^{2}+4\right) K\left( -\frac{4}{x^{2}}\right)
-\left( x^{2}-4\right) E\left( -\frac{4}{x^{2}}\right) \right] ,
\label{eq:g_def}
\end{equation}%
with the complete elliptic integrals of the first and second kinds,
respectively.
\begin{equation}
K(x)=\int_{0}^{1}\frac{\mathrm{d}t}{\sqrt{(1-t^{2})(1-xt^{2})}},\qquad
\qquad E(x)=\int_{0}^{1}\mathrm{d}t\,\sqrt{\frac{1-xt^{2}}{1-t^{2}}}.
\label{eq:Def_elliptic}
\end{equation}%
In this case the normal-frequency mode can be expressed as

\begin{equation}
\omega _{0,\parallel }=\omega _{p}\sqrt{1-2y(x)},  \label{L_spec}
\end{equation}%
where $\omega _{p}=({n_{0}e^{2}}/\varepsilon _{0}{m_{\mathrm{e}}})^{1/2}$,
and the electron density is $n_{0}=N_{\mathrm{e}}/\pi R^{2}L_{n}$. In the
thin-disk limit $(x=L_{n}/R\ll 1)$ we obtain \cite{Downing:2020}
\begin{equation}
\omega _{0,\parallel }^{\mathrm{disk}}\simeq \omega _{\mathrm{p}}\left\{ 1-%
\frac{L_{n}}{4\pi R}\left[ 6\ln {2}-1-2\ln {\left( \frac{L_{n}}{R}\right) }%
\right] \right\} ,\qquad L_{n}/R\ll 1,  \label{w_parall}
\end{equation}%
which approaches the $\omega _{p}$ in the limit of ($L_{n}/R\rightarrow 0$),
see details in Fig. \ref{FigDow}.

The case of the transverse plasmonic mode of the TE type ($\alpha =\perp $)
can be studied in a similar way as done above for the longitudinal mode,
which results in the following expression for the transverse mode frequency
\begin{equation}
\omega _{0,\perp }=\omega _{p}\sqrt{y(x)},~~x\equiv L_{n}/R.  \label{T_dir}
\end{equation}

In Fig.~\ref{FigDow} the solid lines show dependencies of $\omega
_{\parallel }/\omega _{p}$ and $\omega _{\perp }/\omega _{p}$ on $x=L_{n}/R$
in the interval $[0,3]$, while the dotted lines show the corresponding
approximate solutions. It is seen that the latter approximations are valid
for $L_{n}/R<0.9$.
\textcolor{black}{Besides, it is seen Fig.~\ref{FigDow} that the contribution of $E_{\perp}$ is small at $|x| \ll 1$, hence the contribution of $E_{\perp}$ becomes comparable with $E_{\parallel}$ at $x \sim 2$, which corresponds to a cylindrical particle, rather than the ring studied here.}

\section{Resonant emission of nanoemitters in the system with NRs}

In this section, we study the dynamics of the system of Eqs. (\ref%
{MaxwellEqs})-(\ref{EqForN_12}), which combines the Maxwell's equations for
the PP field with the semiclassical approximation for the optical field
emission from the NEs. We apply the FDTD technique to calculate the field in
the NR lattice with the incorporated NEs. Our system (see Fig.~(\ref%
{Fig_Latt_Emitts})) contains the periodic lattice of conducting NRs (of the
size $7\times 7$) with embedded NEs emitting the optical field. PPs are
excited in the NRs and interact with the NEs. In Fig.~(\ref{Fig_Latt_Emitts}%
), the NEs are connected by dashed blue lines that correspond to the TSP
optimized path (when, as said above, a traveling photon passes each NE,
without visiting the same NE twice \cite{BURLAK:2023b}). We use an advanced
technique where the standard FDTD approach is extended by calculating the
dynamics of the semiclassical polarization system (see Eq. (\ref{PolarizEq}%
)) coupled to the population dynamics in the four-level laser NEs (Eqs. (\ref%
{EqForN_03})-(\ref{EqForN_12})), at each time step, see further details in
Ref. \cite{Burlak:2023}.
\textcolor{black}{In this work, numerical calculations were performed by means of the standard FDTD technique \cite{taflove:2005}. Also, the open source package \textbf{fdtd} was used, which allowed us to include the necessary extensions related to the modeling of the Drude frequency dispersion in NRs, see
	link [https://github.com/flaport/fdtd] for further details.}
To simulate the field dynamics in the hybrid system with open boundaries the
standard perfectly matched layer (PML) boundary conditions are applied at
boundaries of the FDTD grid, to avoid the reflection of electromagnetic
waves from the boundary \cite{taflove:2005}. On the adopted scale, the
typical NE\ size is orders of magnitude smaller than the typical NR size,
therefore the NEs are approximated here by the point-like sources. In our
simulations we dealt with the general 3D vector electromagnetic fields $%
\mathbf{E}$ and $\mathbf{H}$ in the general form. The simulations
demonstrate that, in the present system, TM electromagnetic waves with
components $[E_{z},H_{x},H_{y}]$ are mainly generated. This conclusion
agrees with the theory Ref. \cite{Downing:2020}, where it was shown that, in
the case of nano-objects in the disk limit ($L_n/R\ll 1$, where $R$ and $L_n$
are the radius and height, respectively) the main contribution yields the $%
E_{z}$ (alias $E_{\parallel }$) longitudinal field, while the transverse
field is conspicuous at $L_n/R\sim 1$ \cite{Downing:2020}. Thus, in the disk
limit one can obtain $\omega _{0,\parallel }\simeq \omega _{p}$. which
approaches the $\omega _{p}$ value in the considered limit of $%
L_n/R\rightarrow 0$. For the TE mode, the transverse frequency is estimated as
$\omega _{0,\perp }/\omega _{p}\sim \sqrt{L_n/R}$, which is small in the
present case, $L_n/R\rightarrow 0$. To make the following figures clear, only
the $|E_{z}|$ field components are displayed in them. The dynamics of the
present hybrid system significantly depends on frequency $\omega _{p}$ of
the dispersive NRs in the terahertz range. Therefore, in the following we
focus on two cases. $viz$., $\omega _{p}=2.3~$THz and $\omega _{p}=2.3\times
10^{-2}~$THz, in which $\omega _{p}$ differs by two order of magnitude.
\begin{figure}[tbp]
\centering
\includegraphics[
width=0.79\textwidth
]{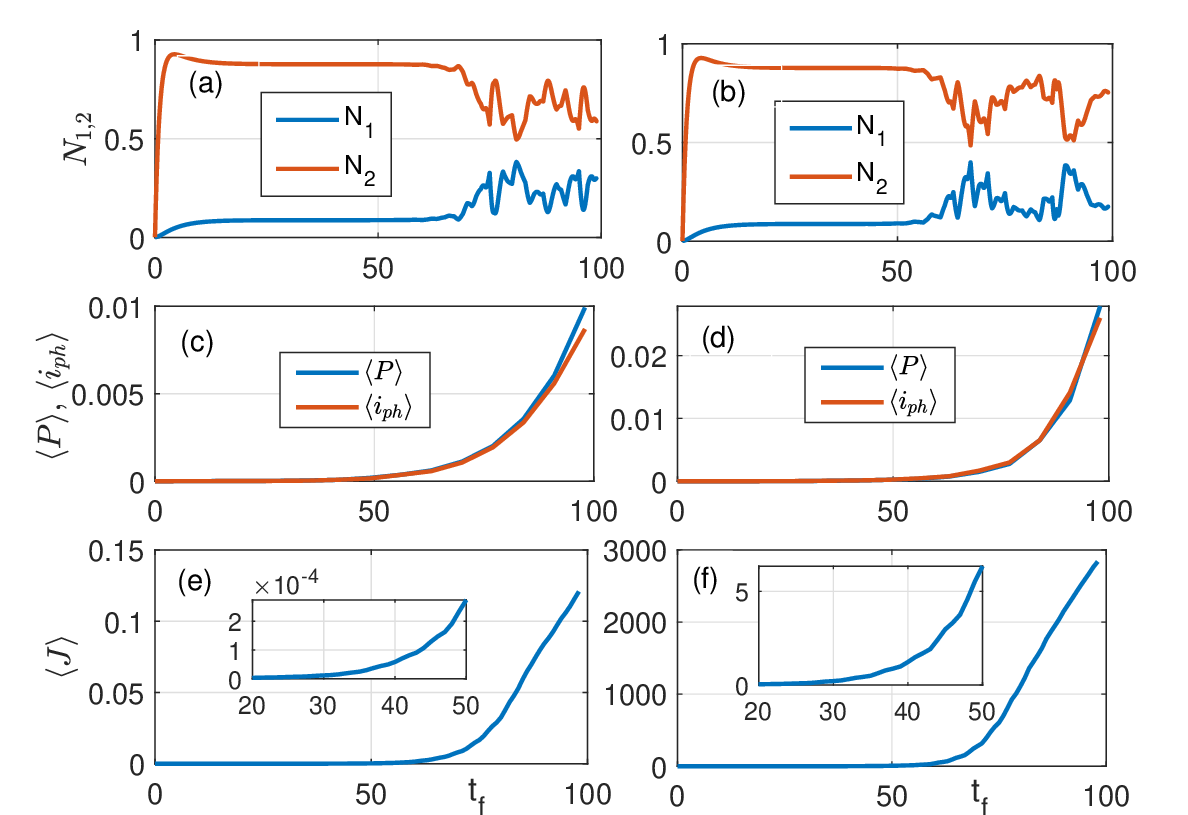}
\caption{The temporal dynamics produced by the numerical solution of the
system of Eq.(\protect\ref{MaxwellEqs})-(\protect\ref{M_rateEq}) for the $%
7\times 7$ NR lattice with two different plasma frequencies: $\protect\omega %
_{p}=2.3\times 10^{-2}~$THz (the left) and $2.3$ THz (the left and right
columns, respectively). Here (a) and (b) show the dynamics of populations $%
N_{1}$ and $N_{2}$ of the NE lasing levels, respectively (see Eq. (\protect
\ref{EqForN_12}); the dynamics of populations $N_{0,3}$, see Eq. (\protect
\ref{EqForN_03}), is not displayed here). (c,d) The average quantum
polarization ${\mathbf{|P|}}$ and photocurrent $i_{ph}={\partial \mathbf{|P|}%
}/{\partial t}$. Panels (e) and (f) exhibit a drastic difference in the
average current $\left\langle J\right\rangle $ (see Eq. (\protect\ref{averJ}%
)) in NRs at $\protect\omega _{p}<\protect\omega _{c}$ (the left) and $%
\protect\omega _{p}>\protect\omega _{c}$ ((e) \ and (f), respectively).}
\label{Fig_3x2_ diff}
\end{figure}
\begin{figure}[tbp]
\centering
\includegraphics[
width=0.68\textwidth
]{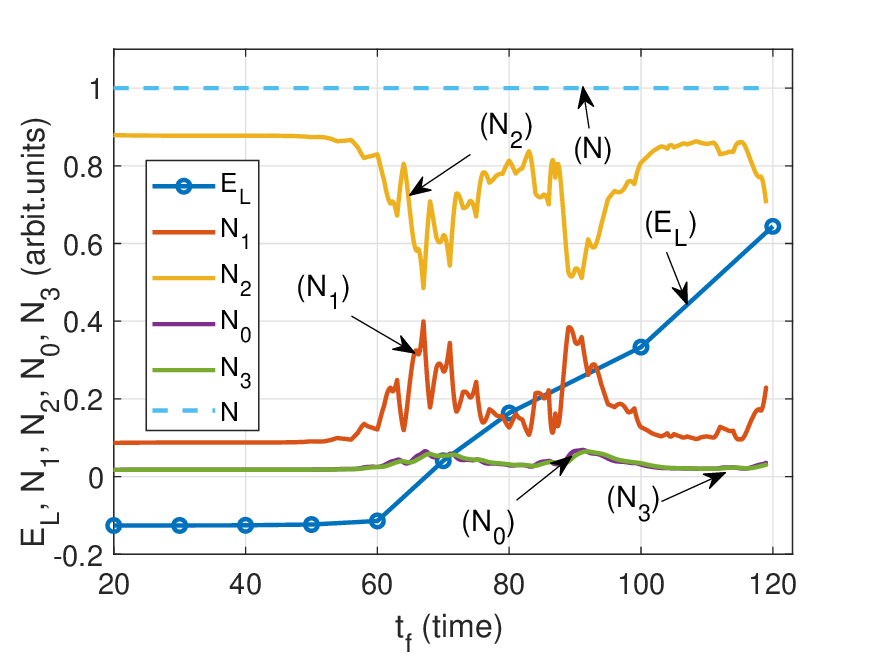}
\caption{The dynamics of populations $N_{0,1,2,3}$ of the $4$-level NE. Red
and yellow lines represent, respectively, populations $N_{1}$ and $N_{2}$ of
the lasing levels. The light blue line shows the conservation of $%
N=N_{0}+N_{1}+N_{2}+N_{3}$. The blue line exhibits the largest Lyapunov
exponent, which changes its value from negative to positive (switching to
instability) at dimensionless calculation time $t_{f}>60$.}
\label{Fig_Popul_Lyap}
\end{figure}

Figure~\ref{Fig_3x2_ diff} displays the dynamics produced by the numerical
solution of Eqs.(\ref{MaxwellEqs})-(\ref{M_rateEq}) for the $7\times 7$ NR
lattice for two different NE plasma frequencies, $\omega _{p}=2.3\times
10^{-2}~$and $2.3$ THz. Panels (a) and (b) display the dynamics of the
populations of the resonant emission NE levels $N_{1,2}$, see Eq.(\ref{EqForN_12}) (the
evolution of $N_{0,3}$ is not displayed here). Panels (c) and (d) show the
evolution of the average quantum polarization ${\mathbf{|P|}}$ and
photocurrent $i_{ph}={\partial \mathbf{|P|}}/{\partial t}$. Further, panels
(e) and (f) show details of the average NR current $\left\langle
J\right\rangle $, see Eq.(\ref{averJ}). We observe that in both cases the
evolution of $\left\langle J\right\rangle $ drastically differs for
different values of $\omega _{p}$. At $\omega _{p}=2.3~$THz, the amplitude
of $\left\langle J\right\rangle $ is significantly larger than at $\omega
_{p}=0.023~$THz. It is seen that, at large times $t>60$, the evolution is
chaotic, weakly depending on details of the NE distribution. Such a behavior
of $N_{1,2}$, displayed in Fig.~\ref{Fig_3x2_ diff}, indicates the onset of
the effective coupling of the PP and NE subsystems, which significantly
depends on $\omega _{p}$.

\begin{figure}[tbp]
\centering
\includegraphics[
width=0.79\textwidth
]{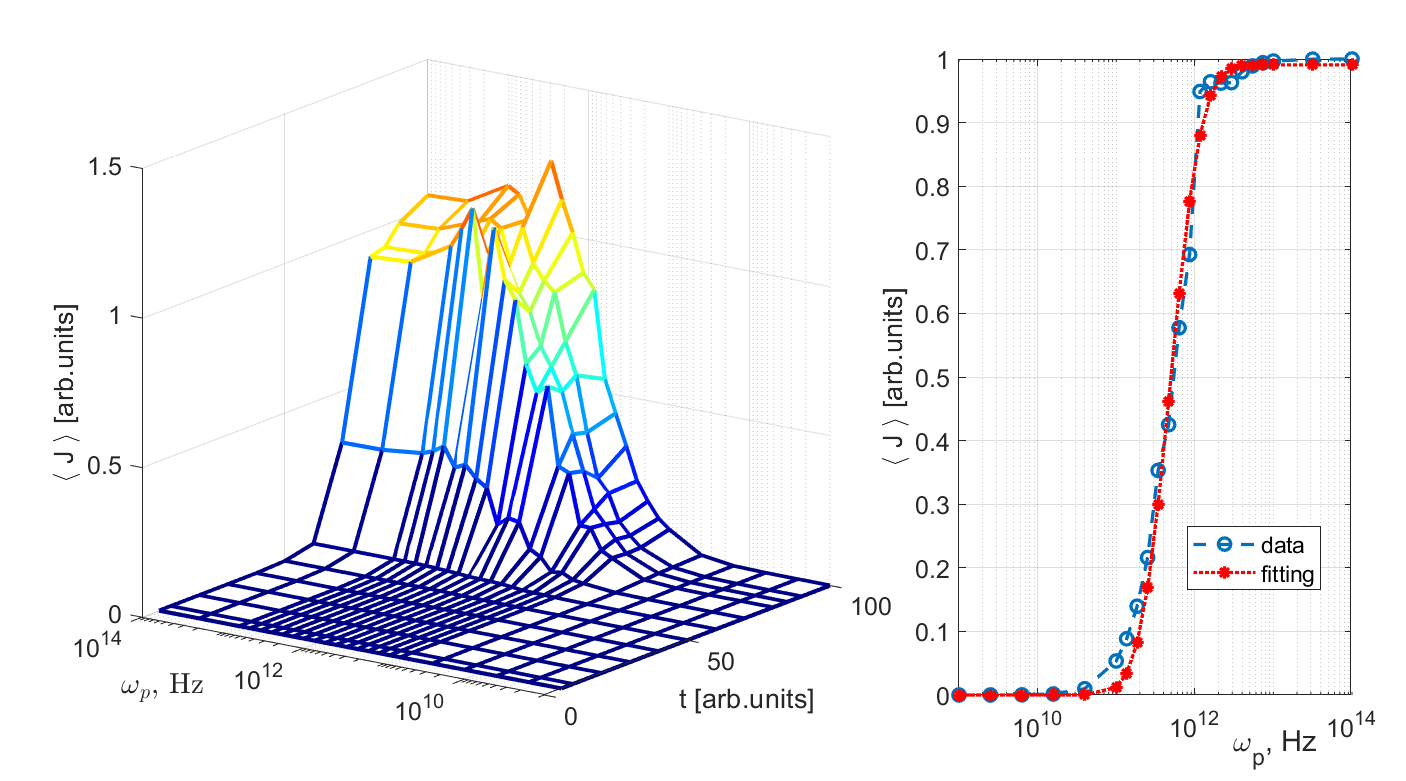}
\caption{(a) The average PP\ current $\left\langle J\right\rangle $ in the
conducting lattice of $7\times 7$ NRs as a function time $t_{f}$ and the NR
plasma frequency $\protect\omega _{p}$. It is seen that $\left\langle
J\right\rangle $ emerges from zero at $t_{f}\approx 55$ for $\protect\omega %
_{p}\geq \protect\omega _{c}\simeq 3\times 10^{11}$ $\mathrm{Hz}$; (b) the
normalized value $\left\langle J(\protect\omega _{p})\right\rangle $ and its
fitting by the $\mathrm{err}(\protect\omega _{p})$ function, which is found
from the FDTD data, see details in the text.}
\label{Fig_AverJ_AproxErrFun}
\end{figure}

\textcolor{black}{Fig.~\ref{Fig_Popul_Lyap} exhibits details of the population dynamics for
	$N_{0,1,2,3}$ of the $4$-level NE at $t_f>20$, when the system commences the transit to the nonlinear regime.
	Red and yellow lines represent, respectively, populations $N_{1}$ and $N_{2}$ of
	the lasing levels, while the green and purple lines correspond to $N_{0}$ and $N_{3}$, respectively. The light blue line shows the conservation of
	$N=N_{0}+N_{1}+N_{2}+N_{3}$. 	
	From  Fig.~\ref{Fig_Popul_Lyap}, it can be seen that, at $t_f > 58$, the stationary values of the population levels become unstable and convert into a nonlinear oscillatory regime. To define the time when the instability commences, we studied the behavior of the Lyapunov function $E_L$. The blue line in Fig.~\ref{Fig_Popul_Lyap}
	instability) at $t_f > 60$.
	 Further analysis shows that such an instability is regularized by the transition to a regime of nonlinear oscillations between the laser levels $N_1$ and $N_2$.}	
To get more insight into the dynamics in the hybrid system, it is instructive to
consider the average current $\left\langle J\right\rangle $ in the NR conductive
lattice, which is
\begin{equation}
\left\langle J\right\rangle =\left\langle J(\omega _{p})\right\rangle
_{N_{r}}=(N_{r})^{-2}\sqrt{\sum \left\vert J_{i,j}\right\vert ^{2}},
\label{averJ}
\end{equation}%
where $J_{i,j}$ is the current in the NR with coordinates $\left( {i,j}%
\right) $, and the summation is performed over the entire NR lattice, $N_{r}$
being the total number of NRs. Figure \ref{Fig_AverJ_AproxErrFun}(a) shows
the average PP current $\left\langle J\right\rangle $ in the conducting
lattice of $7\times 7$ of NRs, see Eq. (\ref{averJ}), as a function of the
scaled simulation time $t_{f}$ and plasma frequency $\omega _{p}$. It is
seen that $\left\langle J\right\rangle $ emerges from zero at $t_{f}\approx
55$ and sharply increases at $\omega _{p}\geq \omega _{c}\simeq 0.5$ THz,
which is a critical value of $\omega _{p}$ indicating the appearance of the
strong coupling between the PP current in the NR lattice and the emission
field in the embedded quantum NEs. The critical value $\omega _{c}$ can be
extracted from the data with the help of the standard fitting technique.
Figure. \ref{Fig_AverJ_AproxErrFun}(b) exhibits the fitting of $\left\langle
J\right\rangle $ by function
\begin{equation}
\left\langle J\left( \frac{\omega _{p}}{\omega _{c}}\right) \right\rangle
\propto F(x)=a\cdot \lbrack \mathrm{erf}(\log \left( x/b\right) )+1],\text{%
where }\mathrm{erf}(z)=\frac{2}{\sqrt{\pi }}\int_{0}^{z}e^{-z^{2}}dz.
\label{averJ_erf}
\end{equation}%
Using the data from our FDTD simulations for $\left\langle J\right\rangle $
and Eq.~(\ref{averJ_erf}) allowed us to extract both fitting parameters $%
a=0.495$ and $b=25$ 
that correspond to $\omega _{c}=4.927\times 10^{11}$ Hz [see the blue and
red lines in Fig.~\ref{Fig_AverJ_AproxErrFun}(b), respectively] within $95\%$
confidence bounds. The numerical package NumPy 2.2.0 (see \textrm{%
https://numpy.org}) was used for this purpose.

To explore details of the nonlinear internal dynamics of the radiating NEs,
we calculated the correlation functions (CF) of the $z$-component of the
photocurrent,
\begin{equation}
I_{ij}(t_{f},\omega _{p})=\left[ \mathbf{I}_{p}\Theta (t_{f})\right] _{z}%
\text{, }||\text{ }z\text{-axis of NRs}  \label{phCurr_z}
\end{equation}%
%
%
%
%
%
%
%
%
%
%
%
%
%
%
%
%
%
%
%
%
for different NE pairs $\left( i,j\right) $ (see Fig.~\ref{Fig_Latt_Emitts})
for various values of $t_{f}$, where $\Theta (x)$ is the Heaviside step
function. The cross-correlation function $\mathrm{CF}(I_{i}I_{j})$ as a
function of the time lag $\tau $ \cite{Therrien:2018, Williams:2013} is ($n$
is NE number)%
\begin{equation}
\mathrm{CF}_{ij}(t_{f},\omega _{p}|\tau )=\mathrm{CF}(I_{i}I_{j})=\frac{1}{{%
ns_{i}s_{j}}}\sum_{m=1}^{n}\Delta I_{i}(t_{m})\Delta I_{j}(t_{m}+\tau )\text{%
,}  \label{CrossCorr}
\end{equation}%
where $\Delta I_{i}\left( t_{m}\right) =I_{i}\left( t_{m}\right) -\overline{I%
}_{i}$, $\overline{I}_{i}$ is the mean value, and $s_{k}^{2}=1/(n-1)%
\sum_{m=1}^{n}\Delta I_{k}\left( t_{m}\right) ^{2}$.

\begin{figure}[tbp]
\centering
\includegraphics[
width=0.68\textwidth
]{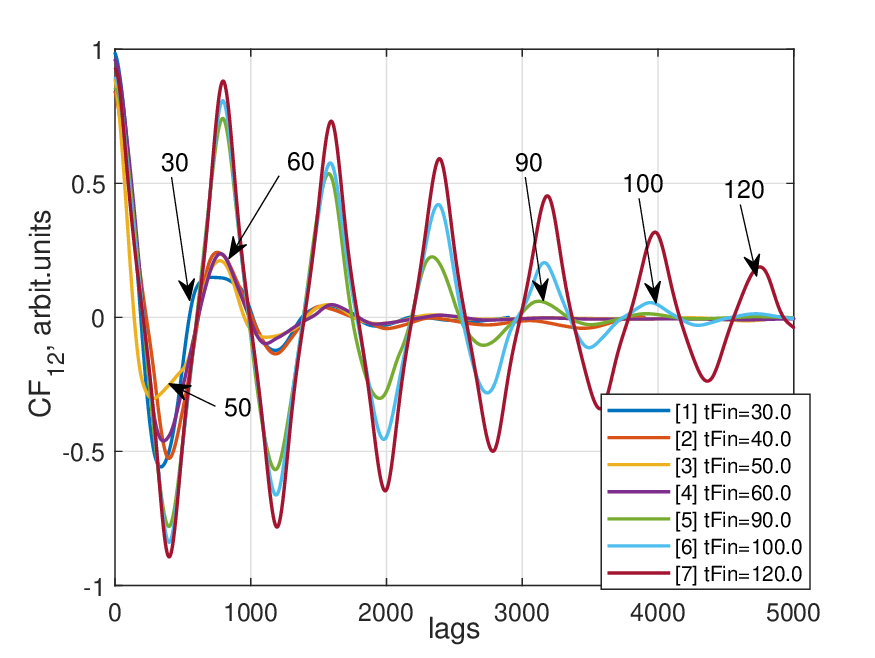}
\caption{The cross-correlation function of the photocurrent $\mathrm{CF}%
_{12} $ for the NEs with numbers $1$ and $2$ at different times $t_{f}$ and $%
\protect\omega _{p}=2.3~$THz, produced by the FDTD solutions of the system
of Eqs.(\protect\ref{MaxwellEqs})-(\protect\ref{M_rateEq}). It is seen that,
at small times, $t_{f}\leq 40$, $\mathrm{CF}_{12}$\ rapidly decays, but at
larger times, $t_{f}\geq 60$, the level of the NE cross-correlations in the
lattice significantly increases.}
\label{Fig_CrossCorrs_vs_lags}
\end{figure}

Figure~\ref{Fig_CrossCorrs_vs_lags} displays the cross-correlation function
of the photocurrents $\mathrm{CF}_{12}$ (see Eq.(\ref{CrossCorr})) for
different times $t_{f}$ at fixed $\omega _{p}=2.3~$THz for the FDTD
solutions of the system of Eqs. (\ref{MaxwellEqs})-(\ref{M_rateEq}). From
Fig.~\ \ref{Fig_CrossCorrs_vs_lags} it is seen that, at small $t_{f}\leq 40$%
, $\mathrm{CF}_{12}$ rapidly decays, but at $t_{f}\geq 60$ the correlations
significantly increase, due to the establishment of collective
synchronization between all the NEs through the laser emission.

\begin{figure}[tbp]
\centering
\includegraphics[
width=1.01\textwidth
]{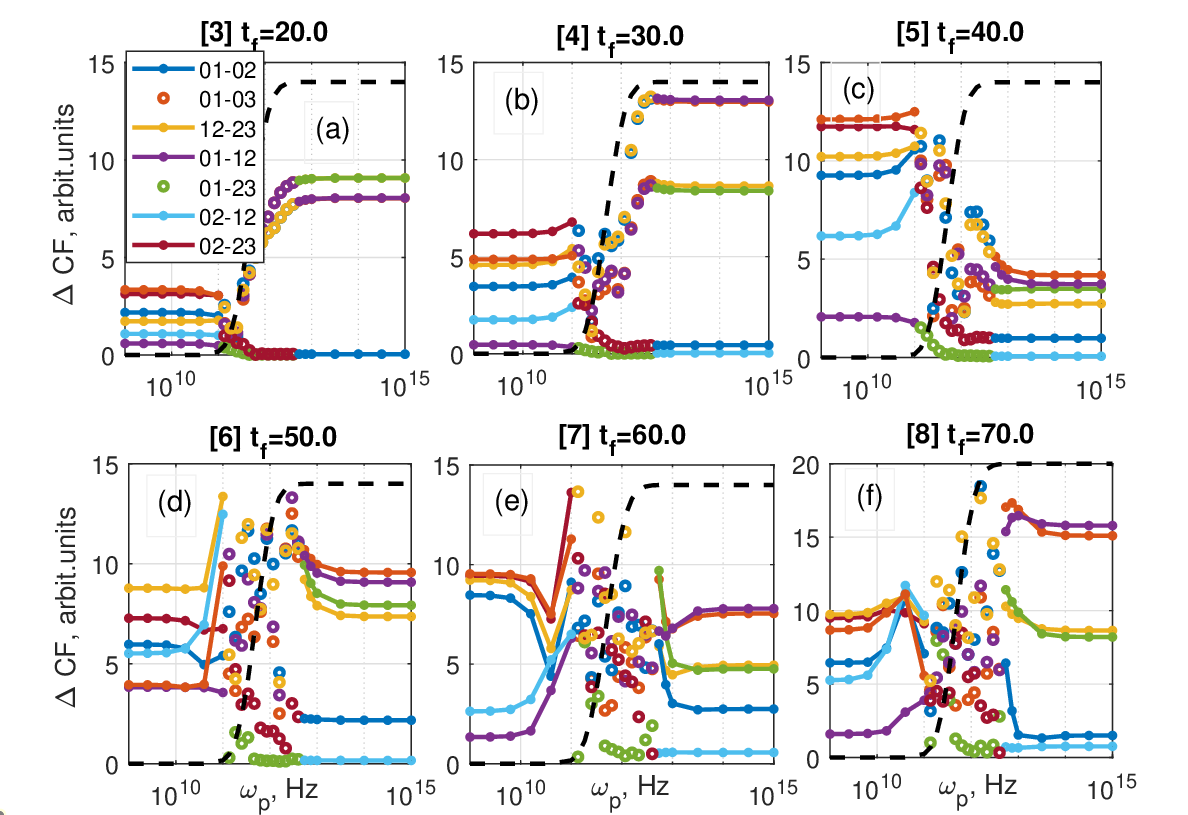}
\caption{The $\widehat{R}(t_{f},\protect\omega _{p})$ norm (see Eq.(\protect
\ref{norm})) of the residual of the cross-correlations of the photocurrents
for different NE pairs $\left( i,j\right) $\ and $\left( k,m\right) $\ (see
the legend in panel (a)) as a function of $\protect\omega _{p}$, for
different $t_{f}$: (a) 20, (b) 30, (c) 40, (d) 50, (e) 60, (f) 70. The color
lines correspond to $\left( i,j\right) -\left( k,l\right) $ NE pairs. The
black line indicates the transition region [see Fig.~(\protect\ref%
{Fig_AverJ_AproxErrFun})(b)] of average current $\left\langle J\right\rangle
$ in the NR lattice [see Eq. (\protect\ref{averJ})]. We observe the
chimera-like states (isolated points) in the transition area.}
\label{Fig_chimeras}
\end{figure}

It is interesting to study the dependence of the correlation on the NR\
plasma frequency $\omega _{p}$. Such a study turns out to be more
instructive when comparing the residuals of the cross-correlation functions
of the NE\ photocurrents,
\begin{equation}
\widehat{R}(t_{f},\omega _{p})=R_{ij,km}(t_{f},\omega _{p})=\sum_{l}\left[
CF_{ij}(t_{f},\omega _{p}|\tau _{l})-CF_{km}(t_{f},\omega _{p}|\tau _{l})%
\right] ^{2},  \label{norm}
\end{equation}%
as a function of $\omega _{p}$. Figure \ref{Fig_chimeras} shows a family of
such dependencies $R(\omega _{p})$ for different times $t_{f}$ and different
NE pairs $(i,j)$\ and $(k,m)$.

In Fig.~\ref{Fig_chimeras}, the black dashed lines indicate the transition
region in terms of the average current $\left\langle J\right\rangle $, see
Eq.(\ref{averJ}). Figure~\ref{Fig_chimeras} demonstrates the appearance, in
the transition region, of a zone similar to chimera states (shown by
isolated circles). In this range, even a small variation of $\omega _{p}$
leads to a significant change in the magnitude of the cross-correlations of
the NE pairs. This fact indicates that, for the coupled PPs and NEs, the
dynamics of the optical field in the quantum NE subsystem essentially
depends on $\omega _{p}$, which is a parameter of the classical NR
subsystem. Further, the black dashed lines allow one to compare the
classical PP\ dynamics in the NRs with the dynamics of the quantum NE\
subsystem (photocurrent) in the critical region, $100$ GHz $<\omega _{p}<1$
THz. From Fig.~\ref{Fig_chimeras} it is seen that, outside the critical
region the branches of the $\widehat{R}(t_{f},\omega _{p})$ curves are
smooth, which makes it possible to connect them by solid lines (as a guide
for the eye). This fact implies a weak coupling between the PP and NE
subsystems in this range, where the conducting NR is actually a dielectric
(roughly, at $\omega _{p}\lesssim 10$ GHz). Note also that, as seen from
Fig.~\ref{Fig_chimeras}, in this configuration $\widehat{R}(\omega _{p})$
remains small for the NE pairs $\left( 0,1\right) -\left( 0,2\right) $ and $%
\left( 0,2\right) -\left( 1,2\right) $, therefore those pairs are
insignificant. This, it is seen in Fig.~\ref{Fig_chimeras}(a) that, at $%
t_{f}=20$, the curve $\widehat{R}(\omega _{p})$ has a characteristic smooth
shape, the details of which depend on $t_{f}$. At $t_{f}=20$, as $\omega _{p}
$ is approaching the critical value $\omega _{c}$ from below, the $\widehat{R%
}(t_{f},\omega _{p})$ remains a smooth function of $\omega _{p}$ with a
local minimum. At $t_{f}=30$, see panel (b), this minimum deepens, and the
dependence $\widehat{R}(t_{f},\omega _{p})$ looses its smoothness in the
critical region. In this zone, due to their chaotic positions, the
connection of adjacent points looses its meaning, and in Fig.~\ref%
{Fig_chimeras} they are appear as a cloud of isolated points. Our
calculations show that, in the critical region, the standard deviation \cite%
{John:2006} of the value of $\widehat{R}(t_{f},\omega _{p})$ is
significantly higher than outside of it. As can be seen from Figs.~\ref%
{Fig_chimeras}(c-f), with the further increase of $\tau _{f}$ towards $%
t_{f}>50$, the apparently chaotic set of points in the critical region cease
changing. Outside the critical region the dependencies $\widehat{R}(\omega
_{p})$ are smooth.


\section{Discussion}

We have addressed the coupling of the NEs (nanoemitters) and PPs (plasma
polaritons) in the hybrid system, built as the periodic lattice of
conducting NRs (nanorings) with embedded NEs, through the common optical
field. The structure of the field significantly depends om on the NR\ plasma
frequency $\omega _{p}$. At the critical value of $\omega _{p}$, the phase
transition occurs in the system, leading to the sharp increase in the
average NR current. In this case, the PP field disturbs the internal degrees
of freedom of the quantum NEs, inducing the cross-correlation of the
photocurrents in all NEs, the correlation magnitude significantly depending
on $\omega _{p}$. The instability of the NE\ resonant emission leads to the appearance
of non-smooth (chimera- or chaos-like) features. In this regime, a small
variation in $\omega _{p}$ leads to a significant change in the magnitude of
the cross-correlations of NE pairs. Figure \ref{Fig_chimeras} exhibits a
possibility of the coexistence of the localized synchronized and
desynchronized cross-correlations of NEs embedded in the lattice of
identical NRs, which results in the formatting of inhomogeneous states in
the hybrid system. Similar (quasi-chaotic) behavior is well known in
Kuramoto systems of coupled oscillators \cite{Davidsen:2024,Lau:2023}.
Patterns featuring this behavior are unstable and are identified as chimeras
\cite{Kemeth:2016}. It is instructive to study the evolution of such
irregular states. Figure~\ref{Fig_NE energy} shows a typical size of the
field regions surrounding PPs in the NR lattice. Initially these areas are
well separated and expand with the increase of time $t_{f}$ up to $t_{f}\sim
80$, when the expanding areas overlap. The latter means that initially
desynchronized (due to the spontaneous emission) NEs develop the
synchronization (induced emission) at $t_{f}\sim 80$. Following the commonly
adopted (scalar) metric $S$ of a curve for the non-smoothness \cite%
{Press:2002}, 
we define the non-smoothness $S$ for our case (the instability of
cross-correlations of NE pairs) as%
\begin{equation}
S(t_{f})=\int \left( \frac{d^{2}f(t_{f},\omega )}{d\omega ^{2}}\right)
^{2}d\omega ,~~f(t_{f},\omega _{p})=\widehat{R}(t_{f},\omega _{p})
\label{Smooth}
\end{equation}%
%
over the curve's domain $\omega _{p}$ (with $\widehat{R}$ defined by Eq. (%
\ref{norm})), see panels (a-f) in Fig.~\ref{Fig_chimeras}.
A lower value of $S(t_{f})$ evidently indicates a smoother curve. (The
natural cubic spline minimizes the $L_{2}$ norm of the second derivative
among all $C_{2}$ the function continuous with its first and second derivatives,
interpolating functions passing through the given points
\cite{Press:2002}.)%
\begin{figure}[tbp]
\centering
\includegraphics[
		width=0.67\textwidth
		]{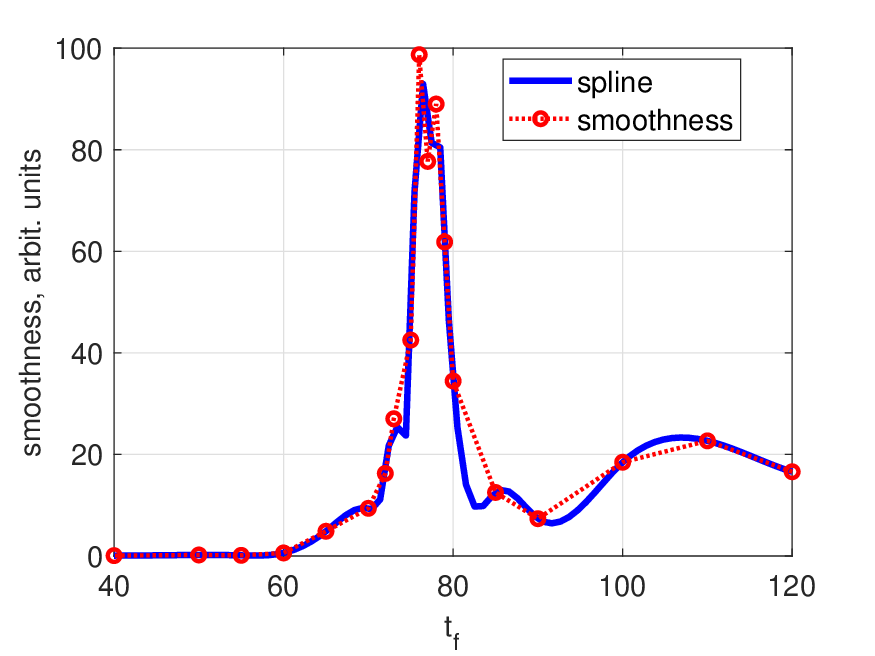}
\caption{The maximum value of the non-smoothness parameter, $S(t_{f})_{\max
}=\max (S_{t_{f}})$ of\ the chimera-like states (red points), displayed in
panels (a-e) of Fig.~\protect\ref{Fig_chimeras}, for all the NE pairs, as a
function of time $t_{f}$ in the area of $\protect\omega \sim \protect\omega %
_{c}$ (see Fig. \protect\ref{Fig_chimeras}). $S_{\max }$ attains the largest
value at $t_{f}\approx 76$, indicating a highly non-smooth structure, as
expected for chimera states \protect\cite{Abrams:2004,Davidsen:2024}.}
\label{ChimBounds1}
\end{figure}
In the present system, the critical behavior occurs in nonlinear quantum NEs
in the respective range of values of the plasma frequency $\omega _{p}$ in
the NR lattice.


As Fig.~\ref{Fig_Popul_Lyap} shows, the maximum Lyapunov
exponent takes positive values in the critical region, indicating unstable
dynamics of the system. Figure~\ref{Fig_NE energy} demonstrates, as said
above, that the effective coupling of the NE oscillators to the PPs in the
periodic NR lattice leads to the dependence of the evolution of the
corresponding chimera-like states on the NR parameter $\omega _{p}$.
\textcolor{black}{Ref.\cite{Song:2016} reported an advanced technology
based on hybrid nanomaterials, such as sandwiched graphene oxide, and
self-assembly of carbon nanotubes (CNT) into CNT rings. The latter, in
particular, allows adjusting $\omega_p$ of CNT rings to the desired range
of $\omega_p$.}

\section{Conclusion}
We have studied the hybrid system, built as the NR lattice carrying PPs
with the nonlinear quantum NEs embedded in the lattice. It is found that the
strong coupling between the PPs and NEs, mediated by the optical field,
leads to the significant cross-correlation between NE pairs, so that the NE
dynamics demonstrates essential dependence on the NR plasma frequency $\omega _{p}$.
However, at larger $\omega _{p} \ge \omega _{c} $ a transition
occurs to the state in which the PP field produces a significant
contribution to the radiation field, which perturbs the dynamics of NEs,
leading to a change in their cross-correlations. In such a system, the NEs
are coupled to the PP field in the NRs, which contributes to the NEs'
optical field. As $\omega _{p}$ increases, the PP field attains a
significant amplitude, which causes a strong coupling between PPs and NEs
and reorientation of the field direction parallel to the PP field in the
NRs. This fact leads to a significant dependence of the shape of the
cross-correlation of photocurrents in NE pairs on the plasma frequency $%
\omega _{p}$ of the NRs coupled to the NEs.


\section{Acknowledgment}
G. M.-A. acknowledges a fellowship provided by the CONAHCYT-M\'{e}xico.\newline

\section{Disclosures}

The authors declare no conflicts of interest.

\section{Data Availability Statement}
The data that support the main findings of this study are available in a publicly accessible repository at the link \url{https://drive.google.com/drive/folders/1w-uolJkl5X0rIBNLozCk85OTC-BbxBl_?usp=sharing} and from the corresponding author upon reasonable request.

\bibliographystyle{unsrt}
\bibliography{MDPI-RefsPhysics1}

@article{Kun:2022,
	title = {Advances in solution-processed quantum dots based hybrid structures for infrared photodetector},
	journal = {Materials Today},
	volume = {58},
	pages = {119-134},
	year = {2022},
	issn = {1369-7021},
	doi = {https://doi.org/10.1016/j.mattod.2022.07.011},
	url = {https://www.sciencedirect.com/science/article/pii/S1369702122001924},
	author = {Kun Ba and Jianlu Wang}
}

@incollection{Perera:2011,
	title = {Chapter 5 - Homo- and Heterojunction Interfacial Workfunction Internal Photo-Emission Detectors from UV to IR},
	editor = {Sarath D. Gunapala and David R. Rhiger and Chennupati Jagadish},
	series = {Semiconductors and Semimetals},
	publisher = {Elsevier},
	volume = {84},
	pages = {243-302},
	year = {2011},
	booktitle = {Advances in Infrared Photodetectors},
	issn = {0080-8784},
	doi = {https://doi.org/10.1016/B978-0-12-381337-4.00005-X},
	url = {https://www.sciencedirect.com/science/article/pii/B978012381337400005X},
	author = {A.G.U. Perera}
}

@article{Ravinder:2020,
	author = {Ravinder K. Jain and Anthony J. Hoffman and Peter Uhd Jepsen and Peter Q Liu and Dmitry Turchinovich and Miriam Serena Vitiello},
	journal = {Opt. Express},
	number = {9},
	pages = {14169--14175},
	publisher = {Optica Publishing Group},
	title = {Mid-infrared, long-wave infrared, and terahertz photonics: introduction},
	volume = {28},
	month = {Apr},
	year = {2020},
	url = {https://opg.optica.org/oe/abstract.cfm?URI=oe-28-9-14169},
	doi = {10.1364/OE.395165},
}

@article{Ryzhii:2020,
	author = {Victor Ryzhii and Maxim Ryzhii and Vladimir Mitin and Michael S. Shur and Taiichi Otsuji},
	journal = {Opt. Express},
	number = {2},
	pages = {2480--2498},
	publisher = {Optica Publishing Group},
	title = {Far-infrared photodetectors based on graphene/black-AsP heterostructures},
	volume = {28},
	month = {Jan},
	year = {2020},
	url = {https://opg.optica.org/oe/abstract.cfm?URI=oe-28-2-2480},
	doi = {10.1364/OE.376299},
}

@article{Arrazola:2014,
	author = {Arrazola, I. and Hillenbrand, R. and Nikitin, A. Yu.},
	title = "{Plasmons in graphene on uniaxial substrates}",
	journal = {Applied Physics Letters},
	volume = {104},
	number = {1},
	pages = {011111},
	year = {2014},
	month = {01},
	issn = {0003-6951},
	doi = {10.1063/1.4860576},
	url = {https://doi.org/10.1063/1.4860576},
	eprint = {https://pubs.aip.org/aip/apl/article-pdf/doi/10.1063/1.4860576/13311002/011111\_1\_online.pdf},
}

@article{Dai:2019,
	author = {Tongyu Dai and Shuangxing Guo and Xiaoming Duan and Renqin Dou and Qingli Zhang},
	journal = {Opt. Express},
	number = {23},
	pages = {34204--34210},
	publisher = {Optica Publishing Group},
	title = {High efficiency single-longitudinal-mode resonantly-pumped Ho:GdTaO4 laser at 2068nm},
	volume = {27},
	month = {Nov},
	year = {2019},
	url = {https://opg.optica.org/oe/abstract.cfm?URI=oe-27-23-34204},
	doi = {10.1364/OE.27.034204},
}

@ARTICLE{Guo:2018,
	AUTHOR = {J. Guo and  K. Black and J. Hu and M. Singh},
	TITLE = {Study of plasmonics in hybrids
	made from a quantum emitter and double metallic nanoshell dimer},
	JOURNAL = {J. Phys. Condens. Matter. },
	VOLUME = {30},
	NUMBER = {},
	PAGES = {, 185301 },
	YEAR = {2018},
	doi = {10.1088/1361-648x/aab72b},
}

@article{Severin:2021a,
	author   = {Habisreutinger, Severin N. and Blackburn, Jeffrey L.},
	title    = {Carbon nanotubes in high-performance perovskite photovoltaics and other emerging optoelectronic applications},
	journal  = {Journal of Applied Physics},
	volume   = {129},
	number   = {1},
	pages    = {010903},
	year     = {2021},
	month    = {01},
	issn     = {0021-8979},
	doi      = {10.1063/5.0035864},
	url      = {https://doi.org/10.1063/5.0035864},
	eprint   = {https://pubs.aip.org/aip/jap/article-pdf/doi/10.1063/5.0035864/20019991/010903\_1\_5.0035864.pdf},
}

@article{Burlak:2024,
	author = {Burlak, Gennadiy and Medina-Angel, Gustavo and Calderon-Segura, Yessica},
	title = "{Plasmon-mediated dynamics and lasing of nanoemitters enhanced by dispersing nanorings}",
	journal = {The Journal of Chemical Physics},
	volume = {161},
	number = {3},
	pages = {034110},
	year = {2024},
	month = {07},
	issn = {0021-9606},
	doi = {10.1063/5.0209350},
	url = {https://doi.org/10.1063/5.0209350},
	eprint = {https://pubs.aip.org/aip/jcp/article-pdf/doi/10.1063/5.0209350/20044198/034110\_1\_5.0209350.pdf},
}

@article{BURLAK:2023b,
	title = {Critical properties of the optical field localization in a three-dimensional percolating system: Theory and experiment},
	journal = {Chaos, Solitons \& Fractals},
	volume = {173},
	pages = {113734},
	year = {2023},
	issn = {0960-0779},
	doi = {https://doi.org/10.1016/j.chaos.2023.113734},
	url = {https://www.sciencedirect.com/science/article/pii/S0960077923006355},
	author = {Gennadiy Burlak and A. Díaz-de-Anda and Boris A. Malomed and E. Martinez-Sánchez and G. Medina-Ángel and R. Morales-Nava and J.J. Martínez-Ocampo and M.E. de-Anda-Reyes and A. Romero-López}
}

@BOOK{Siegman:1986,
	author = {A. E. Siegman},
	year = {1986},
	title = {Lasers},
	publisher = {Mill Valley, California}
}

@article{Burlak:2015,
	author = {Burlak, Gennadiy and Rubo, Y. G.},
	title = {Mirrorless lasing from light emitters in percolating clusters},
	journal = {Phys. Rev. A},
	volume = {92},
	issue = {1},
	pages = {013812},
	numpages = {7},
	year = {2015},
	month = {Jul},
	publisher = {American Physical Society},
	doi = {10.1103/PhysRevA.92.013812},
	url = {https://link.aps.org/doi/10.1103/PhysRevA.92.013812}
}

@Article{Soukoulis:2000,
	author    = {Jiang, Xunya and Soukoulis, C. M.},
	title     = {Time Dependent Theory for Random Lasers},
	journal   = {Phys. Rev. Lett.},
	year      = {2000},
	volume    = {85},
	pages     = {70--73},
	month     = {Jul},
	doi       = {10.1103/PhysRevLett.85.70},
	issue     = {1},
	numpages  = {0},
	publisher = {American Physical Society},
	url       = {https://link.aps.org/doi/10.1103/PhysRevLett.85.70},
}

@BOOK{taflove:2005,
	author = {A. Taflove and S.C. Hagness},
	year = {2005},
	title = {Computational Electrodynamics: The Finite-Difference Time-Domain Methods},
	publisher = {Artech House (Boston)}
}

@article{Downing:2020,
	author = {Downing, Charles A.  and Weick, Guillaume },
	title = {Plasmonic modes in cylindrical nanoparticles and dimers},
	journal = {Proceedings of the Royal Society A: Mathematical, Physical and Engineering Sciences},
	volume = {476},
	number = {2244},
	pages = {20200530},
	year = {2020},
	doi = {10.1098/rspa.2020.0530},
	
	URL = {https://royalsocietypublishing.org/doi/abs/10.1098/rspa.2020.0530},
	eprint = {https://royalsocietypublishing.org/doi/pdf/10.1098/rspa.2020.0530}
}

@Book{Taflove:2005a,
	title     = {Computational Electrodynamics: The Finite-Difference Time-Domain Method, 3rd ed.},
	publisher = {Artech House, Norwood, MA},
	year      = {2005},
	author    = {A. Taflove and S. C. Hagness},
}

@article{Cao_exper:1999,
	title = {Random Laser Action in Semiconductor Powder},
	author = {Cao, H. and Zhao, Y. G. and Ho, S. T. and Seelig, E. W. and Wang, Q. H. and Chang, R. P. H.},
	journal = {Phys. Rev. Lett.},
	volume = {82},
	issue = {11},
	pages = {2278--2281},
	numpages = {0},
	year = {1999},
	month = {Mar},
	publisher = {American Physical Society},
	doi = {10.1103/PhysRevLett.82.2278},
	url = {https://link.aps.org/doi/10.1103/PhysRevLett.82.2278}
}

@article{Tong:2009,
	author = {Tong-Biao Wang and Xie-Wen Wen and Cheng-Ping Yin and He-Zhou Wang},
	journal = {Opt. Express},
	number = {26},
	pages = {24096--24101},
	publisher = {Optica Publishing Group},
	title = {The transmission characteristics of surface plasmon polaritons in ring resonator},
	volume = {17},
	month = {Dec},
	year = {2009},
	url = {https://opg.optica.org/oe/abstract.cfm?URI=oe-17-26-24096},
	doi = {10.1364/OE.17.024096},
}

@article{Wolff:1971,
	author    = {Wolff, I. and Knoppik, N.},
	title     = {Microstrip Ring Resonator and Dispersion Measurement on Microstrip Lines},
	journal   = {Electronics Letters},
	volume    = {7},
	number    = {26},
	pages     = {779--781},
	year      = {1971},
	doi       = {10.1049/el:19710532},
	url       = {https://digital-library.theiet.org/doi/abs/10.1049/el:19710532},
	eprint    = {https://digital-library.theiet.org/doi/pdf/10.1049/el:19710532}
}

@article{Burlak:2023,
	title = {Extended dynamics and lasing of nanoemitters enhanced by dispersing single-walled carbon nanotubes},
	journal = {Journal of Quantitative Spectroscopy and Radiative Transfer},
	volume = {296},
	pages = {108463},
	year = {2023},
	issn = {0022-4073},
	doi = {https://doi.org/10.1016/j.jqsrt.2022.108463},
	url = {https://www.sciencedirect.com/science/article/pii/S0022407322003983},
	author = {Gennadiy Burlak and Gustavo Medina-Ángel}
}

@book{Therrien:2018,
	title={Probability and random processes for electrical and computer engineers},
	author={Therrien, Charles and Tummala, Murali},
	year={2018},
	publisher={CRC press}
}

@article{Williams:2013,
	title = {Experimental Observations of Group Synchrony in a System of Chaotic Optoelectronic Oscillators},
	author = {Williams, Caitlin R. S. and Murphy, Thomas E. and Roy, Rajarshi and Sorrentino, Francesco and Dahms, Thomas and Sch\"oll, Eckehard},
	journal = {Phys. Rev. Lett.},
	volume = {110},
	issue = {6},
	pages = {064104},
	numpages = {5},
	year = {2013},
	month = {Feb},
	publisher = {American Physical Society},
	doi = {10.1103/PhysRevLett.110.064104},
	url = {https://link.aps.org/doi/10.1103/PhysRevLett.110.064104}
}

@book{John:2006,
	title={Probability and random processes for electrical and computer engineers},
	author={John A. Gubner},
	year={2006},
	publisher={Cambridge University Press \& Assessment}
}

@article{Abrams:2004,
	title = {Chimera States for Coupled Oscillators},
	author = {Abrams, Daniel M. and Strogatz, Steven H.},
	journal = {Phys. Rev. Lett.},
	volume = {93},
	issue = {17},
	pages = {174102},
	numpages = {4},
	year = {2004},
	month = {Oct},
	publisher = {American Physical Society},
	doi = {10.1103/PhysRevLett.93.174102},
	url = {https://link.aps.org/doi/10.1103/PhysRevLett.93.174102}
}

@article{Davidsen:2024,
	author = {Davidsen, Jörn and Maistrenko, Yuri and Showalter, Kenneth},
	title = {Introduction to Focus Issue: Chimera states: From theory and experiments to technology and living systems},
	journal = {Chaos: An Interdisciplinary Journal of Nonlinear Science},
	volume = {34},
	number = {12},
	pages = {120402},
	year = {2024},
	month = {12},
	issn = {1054-1500},
	doi = {10.1063/5.0249682},
	url = {https://doi.org/10.1063/5.0249682},
}

@article{Lau:2023,
	title={Chimera patterns in conservative Hamiltonian systems and Bose--Einstein condensates of ultracold atoms},
	author={Lau, Hon Wai Hana and Davidsen, J{\"o}rn and Simon, Christoph},
	journal={Scientific Reports},
	volume={13},
	number={1},
	pages={8590},
	year={2023},
	publisher={Nature Publishing Group UK London}
}

@article{Kemeth:2016,
	author = {Kemeth, Felix P. and Haugland, Sindre W. and Schmidt, Lennart and Kevrekidis, Ioannis G. and Krischer, Katharina},
	title = {A classification scheme for chimera states},
	journal = {Chaos: An Interdisciplinary Journal of Nonlinear Science},
	volume = {26},
	number = {9},
	pages = {094815},
	year = {2016},
	month = {08},
	issn = {1054-1500},
	doi = {10.1063/1.4959804},
	url = {https://doi.org/10.1063/1.4959804},
	eprint = {https://pubs.aip.org/aip/cha/article-pdf/doi/10.1063/1.4959804/14615487/094815\_1\_online.pdf},
}

@Book{Press:2002,
	author = {William H Press and  Saul A Teukolsky and  William T Vettering and  Brian P Flannery},
	title     = {Numerical recipes example book (c++): The art of scientific computing},	
	publisher = {Cambridge University Press},
	address   = {Cambridge}, 
	year      = {2002},	
}

@article{Song:2016,
  author = {Song, J. and Wang, F. and Yang, X. and Ning, B. and Harp, M.G. and Culp, S.H. and Hu, S. and Huang, P. and Nie, L. and Chen, J.},
  title = {Gold Nanoparticle Coated Carbon Nanotube Ring with Enhanced Raman Scattering and Photothermal Conversion Property for Theranostic Applications},
  journal = {Journal of the American Chemical Society},
  year = {2016},
  volume = {138},
  pages = {7005-7015},
  doi = {10.1021/jacs.5b13475}
}

\end{document}